\documentclass[final,1p,times]{elsarticle} 
\usepackage{graphicx} 
\usepackage{amssymb} 
\usepackage{amsthm} 
\usepackage{lineno} 

\journal{Nuclear Physics A} 
\begin{document} 

\begin{frontmatter} 


\title{Energy Loss of Heavy Quarks in a QGP with a Running Coupling Constant Approach}

\author{P.B. Gossiaux  and J. Aichelin}

\address{SUBATECH, UMR 6457, Universit\'e de Nantes, EMN, IN2P3/CNRS
\\ 4 rue Alfred Kastler, 44307 Nantes cedex 3, France}

\begin{abstract} 
We show that the effective running coupling constant, $\alpha_{\rm eff}$, and the effective regulator,
$\kappa \tilde{m}_{D}^2$, which we used recently to calculate the energy loss, $\frac{dE}{dx}$, and the
elliptic flow, $v_2$, of heavy quarks in an expanding  quark gluon plasma plasma (QGP)
\cite{Gossiaux:2008jv} are compatible with lattice results and with recently advanced analytical pQCD
calculation.
\end{abstract} 

\end{frontmatter} 



\section{Introduction}
The analysis of the spectra of heavy mesons,
observed by single non photonic electrons in heavy ion collisions at 200 AGeV \cite{Abelev:2006db,Adare:2006nq},
have revealed two surprising results:
a) Small values of $R_{AA}=d\sigma_{AA}/dp_T^2/(<N_c>d\sigma_{pp}/dp_T^2)$, where
$<N_c>$ is the average number of initial binary collisions
b) A substantial elliptic flow $v_2=<\cos{2(\phi-\phi_{reaction})}>$.
Initially heavy quarks have no elliptic flow and without any interaction we expect $R_{AA}$ to be one
(assuming that the parton
distribution function is similar in pp and in heavy ions).
The finite value of $v_2$ and the low $R_{AA}$ value point therefore
to a substantial interaction with light partons during the
expansion of the system. If in heavy ion collisions a  QGP is formed the natural starting point to calculate the
interaction with the constituents of the plasma, the light quarks and gluons, is perturbative QCD (pQCD).
Comparison with data on $R_{AA}$ and on $v_2$ shows
that the interaction has to be much stronger than obtained in pQCD calculations using a standard fixed value
of the coupling constant $0.3 \le \alpha_s \le à.5$ and a infrared regulator $\mu^2$ in the order of the thermal
squared Debye  mass, $m_D^2 = (1+\frac{N_f}{6})4\pi\alpha_s T^2$, where $N_f$ is the number of flavors.

The weak point of several models aiming at reproducing these $R_{AA}$ and $v_2$ observables is the assumptions on
$\alpha_s$ and on $\mu$, both being not well determined and more or less taken as input parameters. Recently we
advanced an approach in which we related these parameters to physical observables \cite{Gossiaux:2008jv}:

$\bullet$ We employ a running nonperturbative effective coupling constant $\alpha_{\rm eff}(Q^2)$ whose value
remains finite if the Mandelstam variable t approaches 0 by adopting the parametrization of
Ref.~\cite{Dokshitzer:1995qm} for the time-like sector and by truncating the 1-loop renormalized
coupling constant in the space-like sector in order to satisfy the so-called ``universality constrain''
postulated in \cite{Dokshitzer:02}, which leads to
\begin{equation}
\alpha \to \alpha_{\rm eff}(Q^2)=\frac{4\pi}{\beta_0}
\left\{ \begin{array}{ll}
\frac12 - \pi^{-1} {\rm atn}( L_+/\pi )& {\rm for}\:Q^2>0\\
\alpha_{\rm sat} & {\rm for}\:-|Q^2|_{\rm sat}<Q^2<0\\
 L_-^{-1} & {\rm for}\:Q^2<-|Q^2|_{\rm sat}
\end{array}
    \right.
\label{eq: alpha(Q2)}
\end{equation}
with $\beta_0 = 11-\frac23\, N_f$, $N_f=3$, $\alpha_{\rm sat}=1.12$, $|Q^2|_{\rm sat}=0.14~{\rm GeV}^2$ and
$L_\pm = \ln(\pm Q^2/\Lambda^2)$ .

$\bullet$ We extend the usual Braaten-Yuan scheme, applied by Braaten and Thoma \cite{Braaten:1991jj} (BT) to
fixed-$\alpha_s$ QCD, to the case of a running coupling constant, introducing
a semi-hard propagator $\frac{1}{Q^2-\lambda m^2_{D,{\rm eff}}}$, with $m^2_{D,{\rm eff}}(T,Q^2)=
\left(1+\frac{N_f}{6}\right)4\pi\alpha_{\rm eff}(Q^2) T^2$, in the $|Q^2|>|t^\star|$ sector in order
to guarantee a maximal independence of the energy loss $\frac{dE}{dx}$ w.r.t. this unphysical $t^\star$ scale.

$\bullet$ We then define an effective one massive-gluon exchange model (OGE) by replacing the standard
polarization in the t-channel \cite{Weldon:1982aq} $\mu^2= \frac {m_D^2}{3}$,
where $m_D$ is the Debye mass, by an effective one, i.e. $\mu^2=\kappa \tilde{m}_{D}^2$, where
$\tilde{m}_D^2(T)=\frac{N_c}{3} \left(1+\frac{N_f}{6}\right) 4\pi \alpha_s(-\tilde{m}_D(T)^2)\,T^2$
is a self-consistent Debye mass. $\kappa$ is determined by calibrating $\frac{dE}{dx}$ to
the one calculated in our (extended) hard thermal loop
calculation. The details of this procedure is found in the appendix of ref. \cite{Gossiaux:2008jv}
and the details of the OGE models labeled ``model E'' in the text of the same reference.

With these improvements we were able to reproduce the centrality
dependence of the experimental $R_{AA}$ up to a factor of 2-3 as well as the experimental value of $v_2$
with collisional energy loss only. For details we refer to \cite{Gossiaux:2008jv}.

\begin{figure}
\centering
\includegraphics[width=0.32\textwidth]{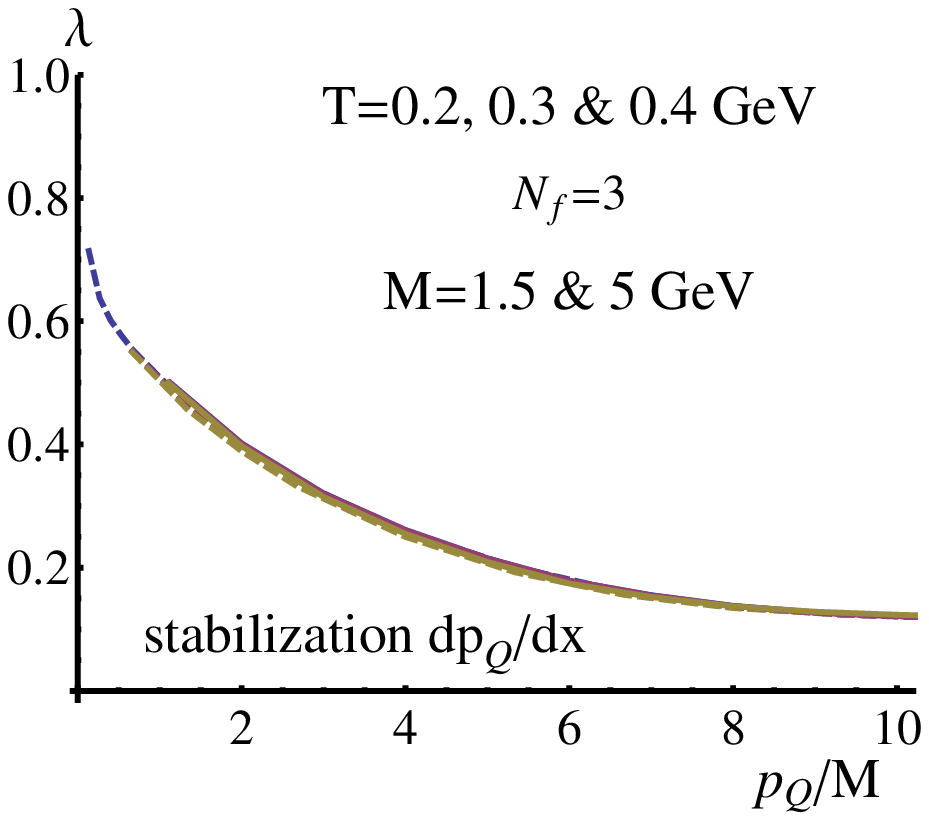}
\includegraphics[width=0.32\textwidth]{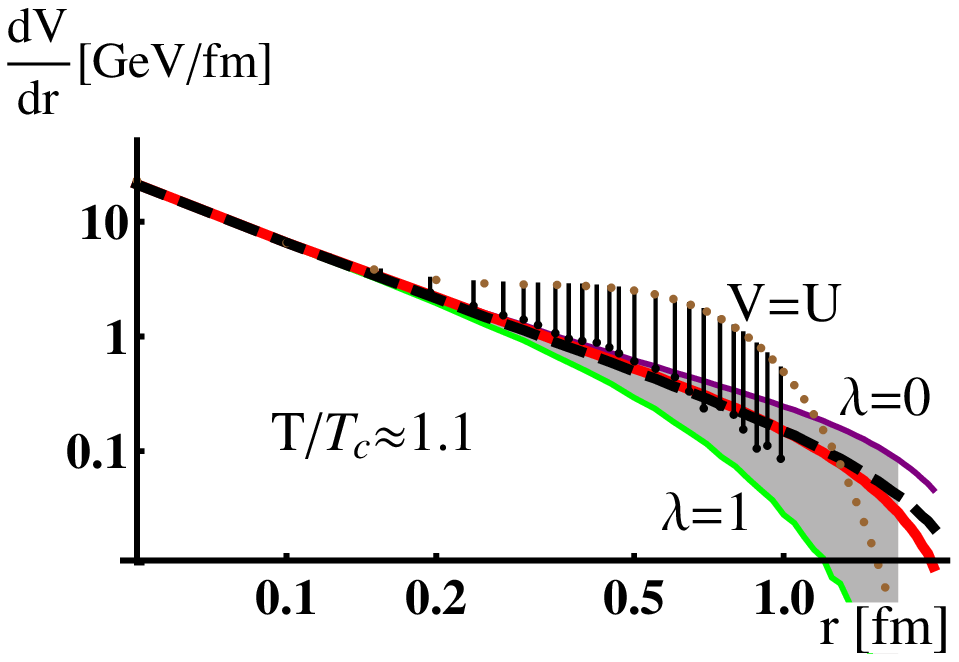}
\includegraphics[width=0.32\textwidth]{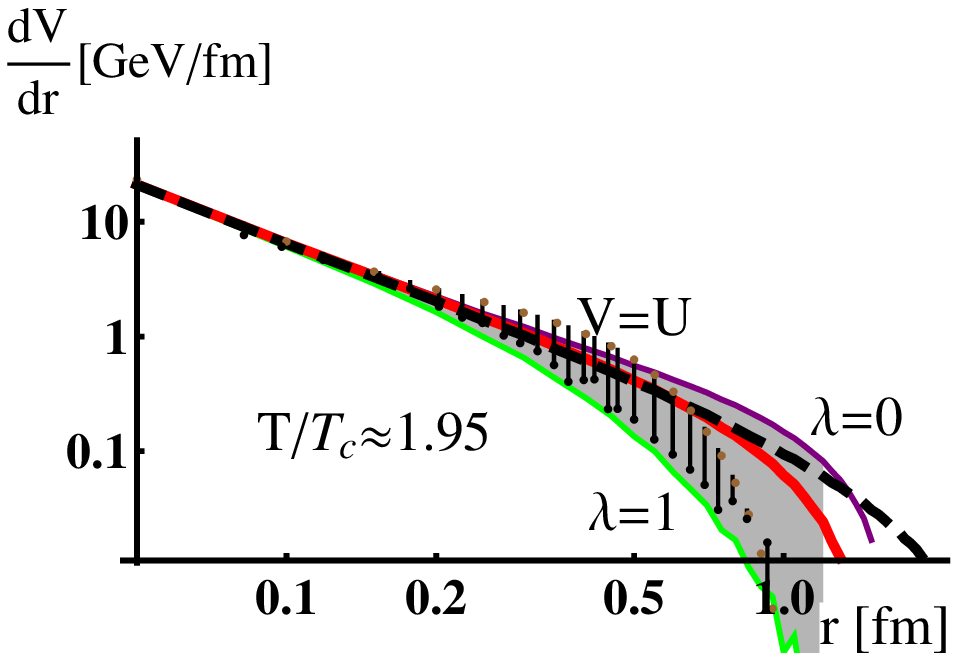}
\caption{(Color online) Left: the $\lambda$ parameter (eq.\ref{htl}) which guarantees the maximal insensitivity of the momentum loss 
per unit length on the unphysical scale $t^*$. Middle and right: Comparison of $dV/dr$ for $\lambda = 0$ and 1 (full curves bordering
the gray bands) and for $\lambda = 0.11$ (thick full red line) and for model E (full dashed black line) of ref. \cite{Gossiaux:2008jv}
with lattice results (vertical lines whose limits correspond to the $V=U$ -- top -- and $V=F$ -- bottom -- prescriptions).}
\label{dau}
\end{figure}

It is the purpose of this contribution to show that our choice of $\alpha_{\rm eff}$ and $\mu$ are
compatible with lattice data. Kaczmarek and Zantow \cite{Kaczmarek:2005ui} have studied
the potential and coupling constant on the lattice in 2-flavor QCD by investigating the free energy
between two heavy quarks.
Assuming that at small distances the singlet potential has the form $V(r)=-\frac{4}{3}\frac{\alpha_s}{r}$
they define $\alpha_{qq}(r,T)=\frac{3}{4} r^2\frac{dV(r)}{dr}$.
In order to compare our approach with these lattice data we start out from the t-channel matrix element,
which consists of two parts:  a hard thermal loop (HTL) approach for ($|Q^2|<|t^*|$) and a Born
``semi hard'' amplitude for ($|Q^2|>|t^*|$)
\begin{eqnarray}
\tilde{V}_{p_Q}(T;\vec{q})&=&
4 \pi \alpha_{\rm eff}(-q^2)\left[
\frac{\Theta(q^2<|t^\star|)}{q^2+\Pi_L(\omega=0,q)}+
\frac{\Theta(q^2>|t^\star|)}{q^2+\lambda(p_Q) m_{D,{\rm eff}}^2(T,-q^2)}
\right]\,,
\label{htl}
\end{eqnarray}
with $q=\|\vec{q}\|$ and $\Pi_L(\omega=0,q)=m^2_{D,{\rm eff}}(T,-q^2)$ and $|t^\star|$ chosen as $T^2$.
Notice that the parameter $\lambda$, and thus our regulator, is found to depend on the heavy
quark momentum $p_Q$ in our prescription, as displayed on figure \ref{dau} (left), but
neither on $T$ nor on the heavy-quark mass $M$ when displayed as a function of
$p_Q/M$\footnote{Standard BT corresponds to
$\lambda=0$, with large numerical dependences of $\frac{dE}{dx}$ on
$t^\star$.}. For $p_Q/M \to 0$, $\lambda$ approaches 1 and our regulator comes into the range of the screening
mass derived for a static potential in lattice QCD \cite{Kaczmarek:2005ui}.

\begin{figure}[ht]
\centering
\includegraphics[width=0.32\textwidth]{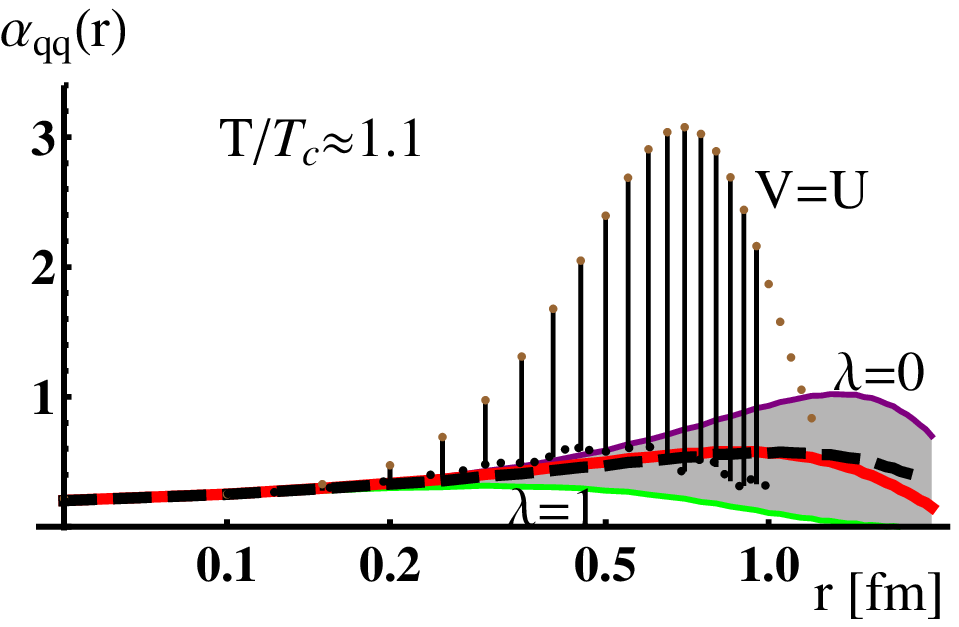}
\includegraphics[width=0.32\textwidth]{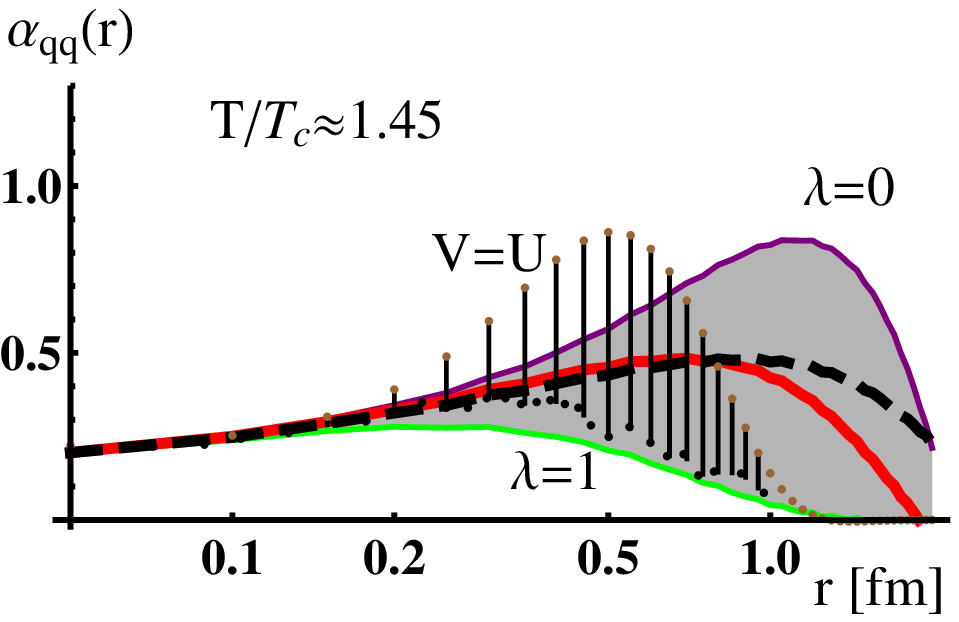}
\includegraphics[width=0.32\textwidth]{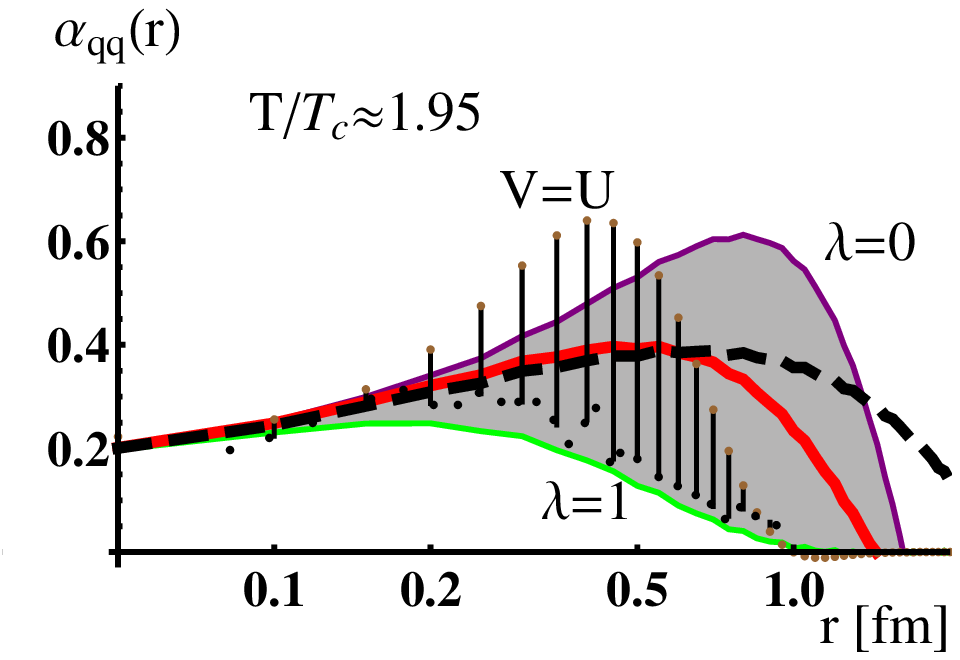}
\caption[]{(Color online) The running coupling constant $\alpha_{qq}$ in our approach as compared to lattice results for 
the same approaches and conventins as in fig.\ref{dau} (middle and right).}
\label{alph}
\end{figure}

The potential in coordinate space is then obtained by Fourier transformation.
The radial force -- with is the essential quantity for $\frac{dE}{dx}$ -- we obtain by derivation. It is
compared in fig.\ref{dau} (middle and right) with results from lattice calculations.
We present our results for three values of $\lambda$: $\lambda$ = 0 (standard BT) and 1 ($\equiv p_Q=0$) present
the boundaries where our approach is meaningful, $\lambda = 0.11$ (red full line) is the value we obtain by the
above procedure for $p_Q \to \infty$.  The lattice results for the free energy F and for the energy U are the
limits of the vertical lines. Whereas at
very small distances as compared to the screening length the potential corresponds to the energy with increasing
distance the polarization of the medium becomes important and the discussion whether the free energy or the energy
presents the Schr\"odinger equivalent potential is not settled yet.
For $p_Q=0$ -- strictly speaking, the only case where the lattice data are relevant for comparison -- our
procedure leads to $\lambda=1$ and a force close to the one deduced for the assumption $V=F$. For increasing
$p_Q$, $\lambda$ decreases correlatively, and our force is in between the two lattice choices $V=F$ and $V=U$ up
to $r\approx 1$ fm. Beyond $r=1~{\rm fm}$, and for $T\approx 2 T_c$ we observe a rather long tail for small
$\lambda$'s and thus relatively moderate values of the associate regulator. The physical relevance of this tail
can only
be addressed once lattice results for $p_Q\gg M$ are available.
An alternative way of comparing our model to lattice data is to resort to the effective T-dependent coupling
constant $\alpha_{qq}$ defined in \cite{Kaczmarek:2005ui}. It is presented in fig.\ref{alph}.

In the interesting regime around 0.5-1 fm, corresponding to 0.2 to 0.4 GeV momentum transfer, the coupling
constant which we employ is around  $T_c$ at the lower limit of that predicted by lattice QCD. Only around
$1.5 T_c$ we start to overpredict $\alpha_s$ for $r>0.8~{\rm fm}$, but this region is of moderate (but finite)
importance for the energy loss of the heavy quark. In figs. \ref{dau} and \ref{alph}, the dashed lines correspond 
to the OGE model E introduced in ref. \cite{Gossiaux:2008jv}. Let us remind that model E was obtained by fixing, for simplicity 
reasons, $\lambda=0.11$ for all values of $p_Q$  and deducing an effective gluon square mass of $\kappa \tilde{m}_D^2$ with 
$\kappa\approx 0.2$.

\begin{figure}
\centering
\includegraphics[width=0.5\textwidth]{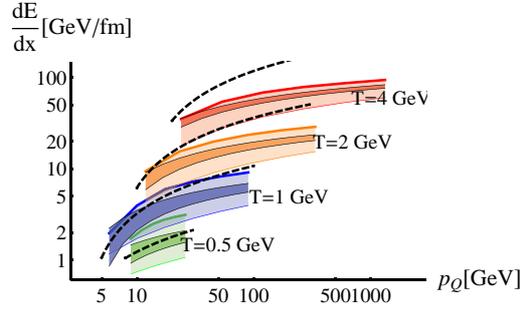}
\caption{(Color online) Comparison of our energy loss with that of ref. \cite{Peigne:2008nd} which uses a running coupling constant calculated in pQCD.
For the symbols we refer to the text.}
\label{pp}
\end{figure}

Our approach can also be compared with a recent analytical approach to derive a running coupling constant by
Peign\'e and Peshier \cite{Peigne:2008nd}. In the limit of large plasma temperatures and therefore of a weak
coupling constant $\alpha_S$ they obtain:
\begin{eqnarray}
-\frac{dE_{\rm coll}}{dx}(E\gg \frac{M^2}{T})&=&\frac{4\pi T^2}{3}\alpha_s(m_D^2)\alpha_s(ET)
\left[\left(1+\frac{N_f}{6}\right) \ln\frac{ET}{m_D^2} \right.\nonumber \\
& + & \left.\frac{2}{9} \ln\frac{ET}{M^2} + c(N_f) +
O\left(\alpha_s(\tilde{m}_D^2) \ln\frac{ET}{\tilde{m}_D^2}\right) \right] \,,
\label{eq_dedx_peshier}
\end{eqnarray}
with $c(N_f)\approx 0.146 N_f +0.05$ and $\tilde{m}_D$ being the self-consistent Debye mass.
Fig.\ref{pp} compares our result with the approach of ref. \cite{Peigne:2008nd}.
The dark intervals mark the results of ref. \cite{Peigne:2008nd}, $\pm \alpha_s(\tilde{m}_D^2)
\ln\frac{ET}{\tilde{m}_D^2}$, whereas our results are the light intervals (bounded by $\lambda = 0.11$ and
$\lambda = 1$). The full line is our result for $\lambda = 0.11$. The dashed line represent the analytical result
for a fixed coupling constant \cite{Peigne:2008nd}. Thus in the range of validity of the approach of ref.
\cite{Peigne:2008nd}, at very high temperatures and/or $p_Q$, our approach reproduces well the most recent
pQCD calculations.

In summary, we have shown that our choice of coupling constant and regulators, motivated by HTL-type calculations,
agrees satisfactorily with lattice and analytical pQCD calculations. The associated drag force deduced in our
approach is an increasing function of the temperature. In particular, it shows no peak around the transition
temperature $T_c$, contrarily to approaches where the $qQ$ interaction is strictly taken as the internal energy
$U$ \cite{VHR:08b}. Phenomenological consequences of these patterns should be addressed systematically in the
future.


\section*{Acknowledgments}
This work was performed under the ANR research program ``hadrons @ LHC'' (grant ANR-08-BLAN-0093-02)
and the PCRD7/I3-HP program TORIC. The authors thank S. Peign\'e and R. Rapp for stimulating discussions.

\end{document}